\documentclass[aps,twocolumn,showpacs,preprintnumbers,pra]{revtex4}
\usepackage{amsfonts}
\usepackage{amsmath}
\usepackage{amssymb}
\usepackage{graphicx}
\usepackage{hyperref}

\newcommand{\bra}[1]{\langle#1|}
\newcommand{\ket}[1]{|#1\rangle}
\newcommand{\eqn}[1]{Eq.~(\ref{#1})}

\begin{document}
%\title{The break-down of the single-shot microwave photon detection via Kerr-type nonlinearity induced by a three-level system}
\title{ Breakdown of the cross-Kerr scheme for Photon Counting}

\author{Bixuan Fan$^{1}$, Anton F. Kockum$^{2}$, Joshua Combes$^{3}$, G\"{o}%
ran Johansson$^{2}$, Io-chun Hoi$^2$, Christopher Wilson$^2$, Per Delsing$^2$, G. J. Milburn$^{1\dag}$ and Thomas M. Stace$^{1\ast}$}
\affiliation{$^1$ Center for Engineered Quantum Systems, School of Mathematics and
Physics, The University of Queensland, St Lucia, Queensland 4072, Australia}
\affiliation{$^{2}$Microtechnology and Nanoscience, Chalmers University of Technology,
S-41296, G\"{o}teborg, Sweden}
\affiliation{$^{3}$Center for Quantum Information and control, University of New Mexico,
Albuquerque, NM 87131-0001, USA. }
\pacs{PACS number}

\begin{abstract}
We show, in the context of single photon detection, that an atomic three-level model for a transmon in a transmission
 line does not support the predictions of the nonlinear polarisability model known as the cross-Kerr effect.  We show that the induced displacement of a probe in the presence or absence of a single photon in the signal field, cannot be resolved above the quantum noise in the probe.  This strongly suggests that cross-Kerr media are not suitable for photon counting or related single photon applications. Our results are presented in the context of a transmon in a one dimensional microwave waveguide, but the conclusions also apply to optical systems.
\end{abstract}

\maketitle

\footnotetext[1]{$^\ast$Corresponding author, stace@physics.uq.edu.au}
\footnotetext[2]{$^\dag$Corresponding author, milburn@physics.uq.edu.au}

The cross-Kerr effect, whereby the phase of one field is changed proportional to the intensity of another, arises from the nonlinear response of an atomic medium to applied fields. It is usually described phenomenologically in terms of a third order term in the nonlinear polarisability, a description that is valid when the applied fields are strong and absorption is weak \cite{Boyd}.  A derivation of the nonlinear polarisability description of the Kerr effect based on an underlying microscopic atomic model has been given by many authors including \cite{DrumWalls,Hilico,Sinclair}.

Many proposed applications of the cross-Kerr effect however suppose that at least one of the fields is very weak --- perhaps only a single photon --- including non-demolition measurements \cite{imoto,grangier1998,munro,roos}, quantum state preparation \cite{ham,gerry,wubiao}, quantum teleportation \cite{vitali} and quantum logic gates build-up \cite{milburn1989,chuang,nemoto,munroJOP,munro05NJP}. All these schemes require strong Kerr nonlinearities at the level of a single photon. It is not clear that the standard model of a cross-Kerr effect, based on a third-order nonlinear polarisability, should be valid for fields with only a few photons.

Doubts regarding the utility of the Kerr effect in single photon applications have been raised before. Shapiro and Razevi \cite{shapiro06,shapiro07} took the multimode nature of the single photon pulse into consideration and found that there is extra phase noise compared to simple single mode calculations, leading to constraints on the achievable phase shifts. Gea-Banacloche \cite{gea10} pointed out that it is impossible to obtain large phase shifts via the Kerr effect with single photon wave-packets. None of this prior work has addressed in detail the question of the cross-Kerr phase shift on a coherent probe field in the presence or absence of a single photon in the control field.

Recently, superconducting circuits have become important test-beds for microwave quantum optics, demonstrating quantised fields,  artificial ``atoms'' (i.e.\ with well-resolved energy levels), and strong ``atom''-field interactions.  The transmon \cite{koch} is one of most promising superconducting artificial atoms due to its insensitivity to  $1/f$ noise, strong anharmonicity, and large dipole moment.  Indeed, the typical size of the transmon is comparable to the dielectric gap in an on-chip microwave waveguide, and  so the dipole moment is within an order of magnitude of the maximum that it can possibly be, given the geometrical constraints of the dielectric gap \cite{devoret2007circuit}. This fact leads to the possibility of very large cross-Kerr nonlinearities, where the transmon provides the non-linear polarisability. Recent experiments using a superconducting transmon in a 1D microwave transmission line have demonstrated gigantic cross-Kerr nonlinearities: a control field with {\em on average} 1 photon induces a phase shift in the probe field of 11 degrees \cite{iochun}. Importantly, in this experiment, the microwave fields were freely propagating; no cavity was involved.

This large cross-Kerr phase shift immediately suggests the possibility of constructing a broadband, number-resolving, microwave-photon counter, as long as the cross-Kerr induced displacement of the probe exceeds the intrinsic quantum noise in the probe. Indeed, broadband microwave photon counting is a crucial missing piece of the experimental quantum microwave toolbox, although there are several proposals for detecting microwave photons \cite{Romero,RomeroPhysica,Johnson,Peropadre,ChenHover}.    %As such,  it is critical to extend models of cross-Kerr non-linearities to the regime in which the ``atomic'' dynamics are explicitly included.

In fact, the cross-Kerr interaction is strictly an effective interaction based on weak field--dipole coupling approximations. Ultimately it is mediated by the strong nonlinearities inherent in an anharmonic oscillator (e.g.\ an atom), so it must eventually break down as a useful description of the physics.  It is therefore important to understand the contribution of the transmon (or atomic) dynamics to the effective nonlinearity in the limit of very strong coupling, which was achieved in \cite{iochun}.   In this work we investigate the coupled field--transmon dynamics in this limit, using proposals for microwave--photon counting as a technical objective to evaluate the validity of the cross-Kerr approximation.

We consider two fields, a probe and a control, incident on a superconducting transmon qubit, which is treated as a three-level, $\Xi$-type system in a one-dimensional transmission line. Such three-level systems are prototypes for analysing cross-Kerr nonlinearities \cite{grangier1998}. We do not eliminate the transmon, but instead treat its dynamics exactly, including quantum noise in the incident fields. The probe is assumed to be a coherent field (or possibly squeezed), while the control field is in a Fock state, whose photon number, $n$, we are trying to measure.  For our purposes, we restrict to $n=0$ or 1.

We show that in spite of the very large cross-Kerr nonlinearity, the induced probe displacement (i.e.\ the signal)  in the presence of a single control photon is limited by saturation effects in the transmon, and is always less than the probe's own quantum noise.  That is, the signal-to-noise ratio (SNR) is always below unity. Moreover, our conclusion also extends to the \textsf{N}-type four-level atomic level configuration, with which cross-Kerr media are often modelled \cite{schmidt,kang,chen,hu}.  These conclusions have rather profound implications for the exploitation of cross-Kerr phenomena in quantum technologies.

\begin{figure}[tb]
\centering
\includegraphics[ width = \columnwidth]{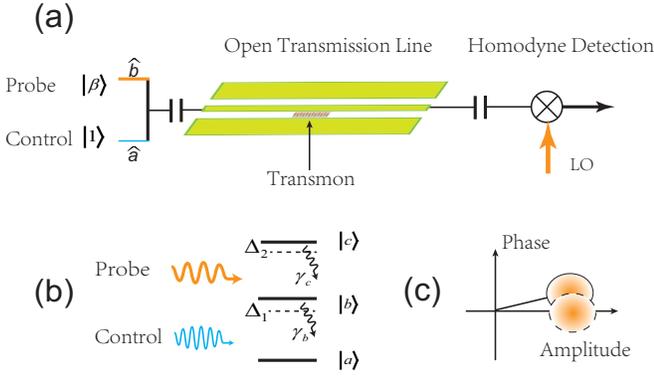}
\caption{{\footnotesize (Color online) (a) Illustrative experimental arrangement. A photon source emits a Fock state microwave photon into  a 1D planar transmission line with a $\Xi$-type three-level transmon embedded in
it. (b) Transmon level structure. The upper transition of the transmon is coupled with a coherent
microwave field (probe) and its lower transition is coupled to the Fock state.  The interaction induced phase shift in the probe field is detected by homodyne detection. (c) Cartoon of the Kerr-induced probe displacement.}}
\label{schematic}
\end{figure}
%
%Three-level models are  convenient for studying
%cross-Kerr nonlinearities \cite{jean}.  The transmon %or the transmission-line shunted plasma oscillation qubit
% is one of most promising superconducting qubits due to its
%insensitivity to  $1/f$ noise, strong anharmonicity, and large dipole moment.  Indeed, the typical size of the transmon is comparable to the dielectric gap in an on-chip microwave waveguide, and  so the dipole moment is within an order of magnitude of the maximum that it can possibly be, given the geometrical constraints of the gap on the field. The model We consider consists of a three-level transmon embedded in a one dimensional transmission line, with a coherent probe and a single photon signal coupled to its upper and lower transitions, respectively. The signal number information is expected to be extracted by measuring the probe phase displacements.

 The transmon levels are $\{\ket{a},\ket{b},\ket{c}\}$, with corresponding energy levels, $\omega_i$, and decay rates, $\gamma_i$, as shown in Fig.\ \ref{schematic}. Relaxation between transmon energy levels is relatively fast compared to dephasing rates, which we neglect. The probe field, $\hat{b}$, is in a coherent state $\ket{\beta}$, and is nearly resonant with the $\ket{b}\leftrightarrow\ket{c}$ transition, whilst the control field is in a Fock state of $n=0$ or 1 photons, at a frequency $\omega_{con}$ close to the $\ket{a}\leftrightarrow\ket{b}$ transition. Qualitatively, the control field induces a transient population transfer into the state $\ket{b}$, and the probe field induces a coherence, $\sigma_{bc}$, between states $\ket{b}$ and $\ket{c}$.  This polarisation couples back to the probe field, so that the  probe field is modified from its input state according to the standard input-output relation
\begin{eqnarray}  \label{inout}
\hat{b}_{out}=\hat{b}_{in}+\sqrt{\gamma_c}\hat{\sigma}_{bc}.\label{inout}
\end{eqnarray}
The homodyne detector monitoring the output probe field yields a photocurrent given by
\begin{eqnarray}
J^{hom}_n(t)=\left\langle \hat{y}\right\rangle+\xi.\label{photo}
\end{eqnarray}
where $\hat{y}=-i\sqrt{\gamma_c}(\hat{\sigma}_{bc}-\hat{\sigma}_{cb})$ is the transmon polarisation, $\hat{\sigma}_{ij}=\ket{i}\bra{j}$ and $%
\xi dt=dW(t)$ is a Weiner process satisfying $E[dW]=0$, $E[d^2W]=dt$.  Finally, the useful signal is the integral of the homodyne current over the lifetime, $T$, of the photon wave packet
\begin{equation}
S_{n}=\int_0^T dt\,J^{hom}_n(t)
\end{equation}
If  $n=0$ the transmon dynamics are trivial, and $E[ S_0]=0$.  For $n=1$, $E[ S_1]\neq0$, and so $S_1$ represents the useful signal associated with a single photon in the control field.  However, in any given measurement, the homodyne current includes quantum noise, characterised by the variance $(\sigma_{S_n})^{2}=E[ S_n^{2}]-E[S_n]^{2}$.   To a good approximation, $\sigma_{S_n}$ is independent of the photon number, $n$, and so we define the signal-to-noise ratio, $\textrm{SNR}=E[ S_1]/(\sqrt{2}\sigma_{S})$.  Note that we assume that the homodyne current  will also include technical noise sources.  We ignore these, so that $SNR$ represents the quantum limit for the proposed scheme.

To study quantitatively the system consisting of a transmon interacting with propagating microwave fields, we adopt two different (but consistent) formulations, yielding both numerical and  analytic results.

In the first formulation we suppose the control photon is generated by a fictitious cavity which is initially in a Fock state.
The field in the cavity decays into the 1D waveguide, and propagates to the transmon, which mediates the interaction between the control and the probe.
We emphasize that the cavity is included simply as a model photon source; the transmon is not contained within the cavity. To analyse this system, we use the stochastic cascaded master equation method \cite{gardiner04,wiseman11}.
%By introducing a stochastic term, the
%conditional dynamics can be captured.
The stochastic master equation describing the conditional dynamics of the cascaded cavity field--transmon density matrix, $\rho$, is given by
\begin{eqnarray}  \label{ME}
d\rho&=& (-i[H_s,\rho]+\gamma_{con}\mathcal{D}[\hat{a}_{con}]\rho +
\mathcal{D}[\hat{L}_b]\rho+\mathcal{D}[\hat{L}_c]\rho)dt  \nonumber\\
&&{}+\sqrt{\gamma_{con}}( [\hat{L}_b,\rho\hat{a}^\dag_{con}]+
[\hat{a}_{con}\rho,\hat{L}_b^\dagger])dt
\nonumber \\
&&{}+\mathcal{H}[\hat{L}_ce^{-i\pi/2}]\rho \,dW\label{me1}
\end{eqnarray}
where $\hat{L}_b=\sqrt{%
\gamma_b}\hat{\sigma}_{ab}$, $\hat{L}_c=\sqrt{\gamma_c}\hat{\sigma}_{bc}$ and
\begin{eqnarray}
H_{s}&=&\Delta _{c}\hat{\sigma}_{cc}+\Delta _{b}\hat{\sigma}_{bb}+\Omega _{p}(%
\hat{\sigma}_{bc}+\hat{\sigma}_{cb}),\\
\mathcal{D}[\hat{r}]\rho &=&\frac{1}{2}(2\hat{r}\rho \hat{r}^{\dag
}-\rho \hat{r}^{\dag }\hat{r}-\hat{r}^{\dag }\hat{r}\rho ),\\
\mathcal{H}[\hat{r}]\rho &=&\hat{r}\rho+\rho\hat{r}^\dag-\rm Tr[\hat{r}\rho+\rho%
\hat{r}^\dag]\rho,
\end{eqnarray}
 $\Delta _{b}=\omega _{ba}-\omega _{con}$, $\Delta _{c}=\Delta
_{p}+\Delta _{b}$ $(\Delta _{p}=\omega _{bc}-\omega _{p})$ and  $\Omega _{p}=\sqrt{\gamma _{con}}\beta $.  We solve \eqn{me1} for the conditional state of the field--transmon system, from which we compute the conditional homodyne photocurrent, using \eqn{photo}.  This approach allows us to generate a simulated measurement record for ensembles of events in which $n=0$ or 1.  From these simulated measurement records, we obtain a histogram of homodyne currents, from which we estimate the SNR.

The second formulation uses the Fock state master equation  \cite{josh,zoller98}. Instead of simulating the free space photon as the output of a fictitious cavity, the propagating photon wave packet drives the transmon directly.  The transmon density matrix acquires indices $m,n$ representing coherences between the transmon and photon Fock subspaces $m$ and $n$. The corresponding master equation for the hierarchy of transmon density matrices $\rho_{m,n}$ is
\begin{eqnarray}
\dot{\rho}_{m,n}(t) &=&-i[H_s,\rho _{m,n}]+\mathcal{D}[\hat{L}_{b}]\rho _{m,n}+%
\mathcal{D}[\hat{L}_{c}]\rho _{m,n} \\\nonumber
&&+\sqrt{n}f^\ast(t)[\hat{L}_{b},\rho _{m,n-1}]+\sqrt{m}f(t)[\rho _{m-1,n},\hat{L}_{b}^{\dagger }]
\notag
\end{eqnarray}%
where  $f(t)$ is a complex valued probability amplitude that determines the photon counting rate as proportional to $|f(t)|^2$.  We first solve the dynamics for $\rho_{0,0}(t)$, which drives  $\rho_{0,1}(t)$ and $\rho_{1,0}(t)$, which in turn drives $\rho_{1,1}(t)$.  We solve these analytically and numerically, and use the quantum regression theorem \cite{lax63} to calculate the SNR (see part A of the supplementary information).  The great advantage of this approach is that we obtain the SNR without resorting to stochastic simulations.

\begin{figure}[tbHp]
 \includegraphics[width = 3.5in]{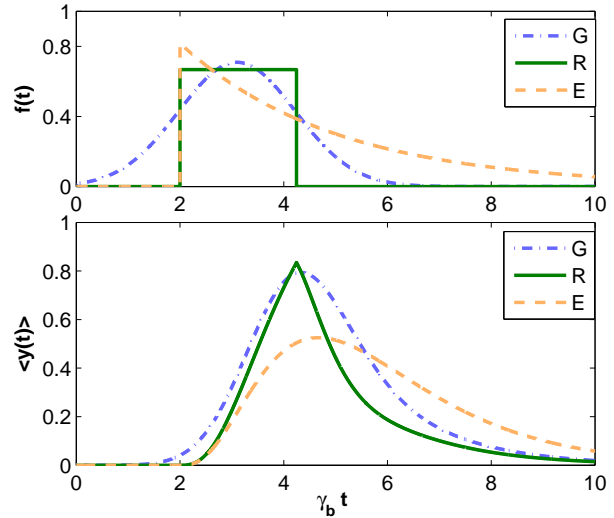}
\caption{\footnotesize (Color online) The transmon responses for different control field wave packets. (top) blue dot-dashed curve:
Gaussian pulse (G); Green solid curve: rectangular pulse (R); Orange dashed
curve: Exponentially-decayed pulse (E); and (bottom)  the corresponding polarisation response of the transmon. The parameters are: $ \Delta_c=\Delta_b=0, \gamma_{con}=0.6672\gamma_b, \gamma_c=2\gamma_b, \beta=0.4\gamma_b (E);0.47\gamma_b(R);0.59\gamma_b(G)$.
}.\label{pulses}
\end{figure}

If the photon is derived from the exponential (E) decay of a cavity mode, then
$f(t)=\sqrt{\gamma _{con}}\exp (-\gamma _{con}t/2)$. Further, this method can handle arbitrary photon wave packets, and we include Gaussian (G) and
rectangular (R), as shown in Fig. \ref{pulses}(top). All pulses contains exactly one photon, that is $\int^T_0 |f(t)|^2 dt=1$ and their common width is $1/\gamma^2_{con}$. The photon induces a polarisation in the transmon, shown in Fig. \ref{pulses}(bottom). We see different pulse shapes yield modest differences in the transmon polarisation $\left\langle\hat{y}(t)\right\rangle$.

%That means the amount of the medium response is not determined by the
%shape, but the total energy of the pulse. Also, it is seen that the response from the transmon (dipole radiation) is always smaller than the input signal in amplitude and the rectangular pulse has slight advantage compared to other two.

% In
%the qutrit-Bloch-like representation the density matrix can be represented
%as $\rho_{m,n} =\frac{1}{3}\mathrm{I} \delta(m,n)+\frac{1}{2}\bar{a}_{m,n}.\bar{\lambda}$
%with $\mathrm{I}$, $\lambda_i (i=1,2...8)$ and $\bar{a}_{m,n}$ being a 3 by 3 identity matrix, the eight Gell-Mann matrices \cite{gells} and their coefficient vector, respectively. Then we solve these equations from the lowest one $\rho_{0,0}(t)$ and
%substitute into the next one $\rho_{0,1}(t)$ and finally solve $\rho_{1,1}(t)$.
%The analytical expressions of ***THe notation $ai11$ is really bad**** $ai11(i=1,2...8)$ and $\rho_{1,1}$ are provided in the supplementary information for the on resonance case.
%
%Experimentally, what really matters is the SNR, not the pure phase
%displacement. For the stochastic master equation method, both mean values of signals and noises can be directly extracted from the conditional detection records. In the second formulation, instead of stochastic simulations, the noise or the variance of the detected signal, can be calculated by the quantum regression theorem \cite{lax}:
%\begin{eqnarray}
%&&(\Delta S)^{2}=E\left[ S^{2}-\bar{S}^{2}\right]  \\\nonumber
%&&=\int^{T}_0 dt\int^{T}_0 dt' u(t'-t) Tr[\hat{y}(t)e^{\mathcal{L}(t'-t)}(-i\hat{L}_2\rho(t)\\\nonumber
%&&+i\rho(t)\hat{L}^\dag_2)]+u(t-t')Tr[\hat{y}e^{\mathcal(t-t')}(-i\hat{L}_2\rho(t')\\\nonumber
%&&+i\rho(t')\hat{L}^\dag_2)]+T-\bar{S}^{2}
%\end{eqnarray}%
%where $S=\int^T_0 dtJ_{hom}(t)$, $\bar{S}=\int^T_0 dt y_{uc}(t)$ and the function $u(t)=1 (t>0); u(t)=0 (t<0)$. The subscripts $uc$ means unconditional results.

\begin{figure}[tb]
\includegraphics[width =3.5in]{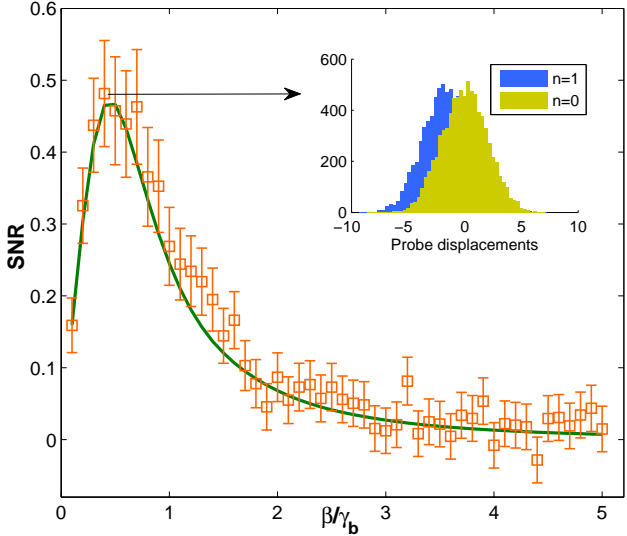}
\caption{\footnotesize (Color online) The SNR as a function of the probe amplitude $\beta$ at optimal parameter setting. The orange
square represents the numerical SNR from the stochastic, cascaded master equation
method and the green curve represents the analytical SNR from the Fock
state master equation method and quantum regression theorem. The inset provides the histogram of the highest SNR for zero and one signal photon.}\label{SNR}
\end{figure}

In Fig. \ref{SNR}, the SNR is shown as a
function of the probe amplitude with other parameters (the detunings and the single photon pulse width) optimised.  The points are calculated by averaging 5000 trajectories of the stochastic master equation, whilst the solid line is computed from the Fock state master equation.  There is good agreement between the two approaches. The inset shows a histogram of the results of the stochastic simulation at the value of $\beta$ that optimises the SNR. Clearly the SNR is everywhere less than unity, and so it is impossible to reliably distinguish between zero and one photon in a single shot. The histogram confirms that the distribution of integrated homodyne current is much broader than the  separation of the means.

% In terms of using a strong probe to compensate low nonlinearity, we can see that over certain intensity value, the SNR gets saturated and no longer increases as the probe amplitude increases but decreases. In fact, At extremely strong probe, the radiation of
%1-2 transition will be largely suppressed and so is the SNR. Therefore, using a strong field or an amplitude-tunable probe field can not overcome the low SNR problem.

\begin{figure}[Htbp]
\includegraphics[width = 3.5in]{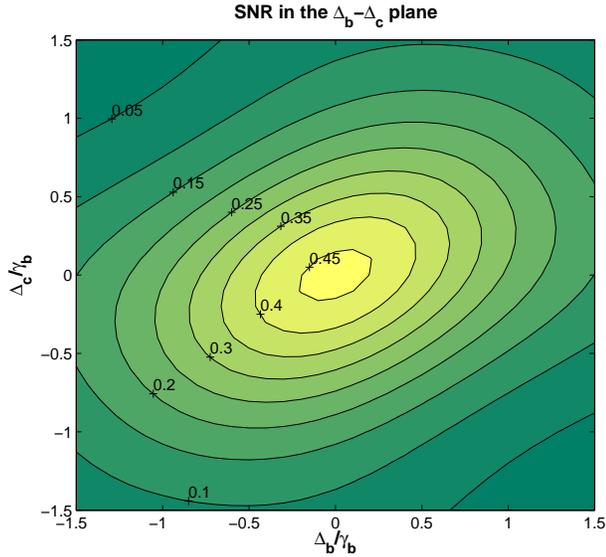}
\caption{\footnotesize (Color online) The SNR of the probe field as functions of detunings
$\Delta_b$ and $\Delta_c$. The other parameters are: $\gamma_c=2\gamma_b$, $\gamma_{con}=0.6772\gamma_b$, $\beta=0.4\gamma_b$.}
\label{detunings}
\end{figure}

As an example of the influence of the various parameters on the SNR, we plot the SNR as a function of the detunings $\Delta_b$ and $\Delta_c$, as illustrated in Fig. \ref{detunings}. Clearly, the optimal SNR is located at $\Delta_b=\Delta_c=0$.

The reason that SNR$<1$ can be understood in the following way: a single control photon induces a variation in the transmon polarisation $\hat \sigma_{bc}$, which manifests as a fluctuation in the homodyne current according to \eqn{photo}.  However the polarisation of the transmon is a bounded operator: $||\hat y||\leq \sqrt\gamma_b$.  The optimal photon wave packet width is $T\sim\gamma_b^{-1}$ (any shorter and the transmon cannot respond to the field; any longer and vacuum noise in the homodyne signal grows), so the integrated polarisation can be no larger than $|E[S_1]|\leq\int_0^Tdt\,||\hat y||\leq\gamma_b^{-1/2}$.  Quantum noise in \eqn{photo} gives $\sigma_{S}^2\geq\textrm{var}[\int_0^Tdt\,\xi]=\gamma_b^{-1}$ we see that the signal to noise ratio is necessarily less than unity.  Fig. \ref{SNR} bears out this analysis: for small probe field amplitudes, the SNR increases quickly, however the transmon dynamics quickly saturates at higher probe amplitudes.

This argument suggests that the fundamental problem is the saturation of the transmon transition.  It may be thought that this can addressed by increasing the number of transmons. We therefore briefly consider a system of $N$ transmons, arranged such that the spacing between adjacent transmons
is much smaller than the wavelength, the transmons are described by the collective atomic spin operators
\begin{eqnarray}
\hat{S}_{ij}=\frac{1}{\sqrt{N}}\sum_k\sigma^k_{ij}
\end{eqnarray}
The stochastic master equation describing
the $n$-transmon system is given by
\begin{eqnarray}
d\rho&=& -i[H_s,\rho]dt+\gamma_{con}\mathcal{D}[\hat{a}_{con}]\rho dt+N\gamma_b%
\mathcal{D}[\hat{S}_{ab}]\rho dt\nonumber \\
&+&N\gamma_c\mathcal{D}[\hat{S}_{bc}]\rho dt-\sqrt{N\gamma_{con}\gamma_b}([\hat{S}%
_{ba},\hat{a}_c\rho]+[\rho\hat{a}^\dag_c,\hat{S}_{ab}])dt  \nonumber \\
&+&\sqrt{N\gamma_c}\mathcal{H}[\hat{S}_{bc}e^{-i\pi/2}]\rho dW  \label{ntrans}
\end{eqnarray}
where
\begin{eqnarray}
H_s=N(\Delta_{c}\hat{S}_{cc}+\Delta_{b}\hat{S}_{b})+\sqrt{N\gamma_c}\beta(\hat{S}%
_{bc}+\hat{S}_{cb})
\end{eqnarray}
We can see that the ensemble master equation (\ref{ntrans}) is the same form as the single-transmon master equation, albeit with decay rates and energies  scaled by a factor of $N$, and its dynamics must therefore be correspondingly faster.  However, this cannot change the optimised SNR, so the SNR for the $N$ transmon case will be the same as for the single transmon case.
%The collective effect makes the interaction happen in a shorter time
%with a higher peak. However, the total amount of transmon radiation is
%almost the same as that in the single transmon case, due to the limitation
%of the response from a single signal photon.

It is worth commenting on a number of other avenues that we have explored, but which yield similar negative results.  Firstly, it may be thought that squeezing the probe field in an appropriate quadrature would reduce the noise, and therefore improve the SNR.  Since we are monitoring the displacement of the probe field, we should squeeze the phase quadrature. However this will enhance noise in the conjugate, amplitude quadrature.  The additional noise in the probe amplitude adds noise to the transmon dynamics arising from fluctuations in $\Omega_p$.  We find numerically that these tradeoffs yield no net improvement in the SNR. (see part B of the supplementary information)

Secondly, there is another multi-transmon limit, in which the transmons are sufficiently separated that they may be considered as a series of cascaded systems.  It might be hoped that the control field will interact sequentially with each transmon.  A local probe at each transmon would then yield an independent estimate of the probe displacement.  After $M$ such independent probes, the SNR would be improved by a factor of $M^{1/2}$, and the SNR could be made arbitrarily large by increasing $M$.   However, the Kramers-Kronig relations impose a  tradeoff between the phase shift at each transmon and the probability that the control field is reflected: a large phase shift necessarily implies a large reflection probability.  Again, we find numerically that the tradeoff yields no net improvement (see part C of the supplementary information).

Thirdly, a number of proposals for inducing cross-Kerr nonlinearities in optical systems use an \textsf{N}-type four-level system  \cite{schmidt,kang}, with a strong classical field addressing the intermediate transition.  As we show in part D of the supplementary information, in the limit of strong driving, this maps onto the same three-level structure we consider in this work, so the conclusions we have reached here also apply to such \textsf{N}-type systems.

Fourthly, We also numerically investigated the effect of varying the ratio $\gamma_c /\gamma_b$ and for $1 < \gamma_c/ \gamma_b < 100$ find that the SNR is still much less than unity (see part E of the supplementary information).

A number of proposals suggest using weak Kerr media to build controlled phase and C-NOT gates with fewer resources than linear optical schemes \cite{nemoto,munro05NJP}. In these schemes the cross-Kerr phase shift per photon is much less than $\pi$, so a strong coherent bus  compensates for the weak nonlinearity, such that the small cross-Kerr phase shift manifests as a large displacement of the strong coherent field.
 However the saturation of the cross-Kerr effect described above indicates that once the displacement of the strong coherent field approaches its own quantum noise, saturation effects lead to the breakdown of the effective cross-Kerr description,  rendering such protocols ineffective.

In summary, we have investigated the feasibility of microwave photon-counting based on an induced cross-Kerr nonlinearity arising from coupling to a large anharmonic dipole. %cross-Kerr- nonlinearity provided by a ladder-type three level system through two different theoretical formulations.
We find that saturation of the transmon transition limits the SNR to less than unity.  As such, it is not possible to use  strong, atom-induced cross-Kerr nonlinearities to perform single photon detection. This conclusion applies to a number of extensions of the basic model, including multiple transmons, cascaded transmons and $\textsf{N}$-type, four-level system.  Further, it limits the applicability of any proposal that requires a cross-Kerr nonlinearity to produce a displacement of a coherent field by an amount greater than the intrinsic quantum noise in the coherent field: it is precisely this condition where the effective cross-Kerr description breaks down, and saturation effects become dominant.

This work is sponsored by ARC Center of Excellence for Engineered Quantum Systems and the China Scholarship council. Bixuan Fan would like to thank Dr.Zhenglu Duan and Dr. Matt Woolley for helpful discussions.

\section*{SUPPLEMENTARY INFORMATION}
%\begin{multicols*}{1}
\subsection{The analytical solution for a three-level system on
resonance}

Here we provide the details of the analytical solution for a three level system. As presented in the main text, we consider a three-level transmon coupling with a coherent field at the $\ket{b}\leftrightarrow\ket{c}$ transition and a single photon at the $\ket{a}\leftrightarrow\ket{b}$ transition. The one-photon Fock state master equations \cite{josh} are given by
\begin{eqnarray}
\dot{\rho}_{0,0}(t) &=&-i[H_s,\rho _{0,0}]+\mathcal{D}[\hat{L}_{b}]\rho _{0,0}+%
\mathcal{D}[\hat{L}_{c}]\rho _{0,0} \\
\dot{\rho}_{0,1}(t) &=&-i[H_s,\rho _{0,1}]+\mathcal{D}[\hat{L}_{b}]\rho _{0,1}+%
\mathcal{D}[\hat{L}_{c}]\rho _{0,1}\\\nonumber
&+&f ^{\ast }(t)[\hat{L}_{b},\rho _{0,0}] \\
\dot{\rho}_{1,1}(t) &=&-i[H_s,\rho _{1,1}]+\mathcal{D}[\hat{L}_{b}]\rho _{1,1}+%
\mathcal{D}[\hat{L}_{c}]\rho _{1,1}\\\nonumber
&+&f^\ast (t)[\hat{L}_{b},\rho _{1,0}]+f (t)[\rho _{0,1},\hat{L}_{b}^{\dagger }]
\end{eqnarray}%
where the temporal profile function $f(t)$, system operators $\hat{L}_b$, $\hat{L}_c$ and the system Hamiltonian $H_s$ have been defined in the main text. Here $\rho_{m,n}$ is a matrix and the subscripts $m$ and $n$ denote the photon number basis.

Initially the transmon is prepared at the ground state. The lowest equation for $\rho _{0,0}(t)$ can be easily solved as $\rho _{0,0}(t)=\rho _{0,0}(0)$. Then it is substituted to the next equation for $\rho_{0,1}(t)$, which is traceless.
For our system and an arbitrary input Fock state, the generalized density matrices $\rho_{m,n}$ in the Bloch-like representation can be parameterized as
\begin{equation}
\rho_{m,n} =\frac{1}{3}\mathrm{I} \delta(m,n)+\frac{1}{2}\bar{a}_{m,n}\bar{\lambda}
\end{equation}%
where I is a 3 by 3 identity matrix and vectors $\bar{a}_{m,n}=(a_{1mn},a_{2mn},a_{3mn},a_{4mn},a_{5mn},a_{6mn},a_{7mn},a_{8mn})$ and
$\bar{\lambda}=(\lambda _{1},\lambda _{2},\lambda _{3},\lambda _{4},\lambda
_{5},\lambda _{6},\lambda _{7},\lambda _{8})$. The Gell-Mann matrices \cite{hioe}  for qutrit are
\begin{equation*}
\lambda _{1}=%
\begin{pmatrix}
0 & 1 & 0 \\
1 & 0 & 0 \\
0 & 0 & 0%
\end{pmatrix}%
\lambda _{2}=%
\begin{pmatrix}
0 & -i & 0 \\
i & 0 & 0 \\
0 & 0 & 0%
\end{pmatrix}%
\lambda _{3}=%
\begin{pmatrix}
1 & 0 & 0 \\
0 & -1 & 0 \\
0 & 0 & 0%
\end{pmatrix}%
\end{equation*}%
\begin{equation*}
\lambda _{4}=%
\begin{pmatrix}
0 & 0 & 1 \\
0 & 0 & 0 \\
1 & 0 & 0%
\end{pmatrix}%
\lambda _{5}=%
\begin{pmatrix}
0 & 0 & -i \\
0 & 0 & 0 \\
i & 0 & 0%
\end{pmatrix}%
\lambda _{6}=%
\begin{pmatrix}
0 & 0 & 0 \\
0 & 0 & 1 \\
0 & 1 & 0%
\end{pmatrix}%
\end{equation*}
\begin{equation}
\lambda _{7}=%
\begin{pmatrix}
0 & 0 & 0 \\
0 & 0 & -i \\
0 & i & 0%
\end{pmatrix}%
\lambda _{8}=\frac{1}{\sqrt{3}}%
\begin{pmatrix}
1 & 0 & 0 \\
0 & 1 & 0 \\
0 & 0 & -2%
\end{pmatrix}%
\end{equation}%
These matrices are traceless, Hermitian, and obey the relation $\mathrm{tr}%
(\lambda _{i}\lambda _{j})=2\delta _{ij}$.

Substituting $\rho_{0,1}$ in terms of $\bar{a}_{0,1}$ into the Fock master equation, we can have the coefficient equations for $a_{i01}(t) (i=1,2,...8)$.  With initial conditions $a_{i01}(0)(i=1,2...8)=0$, we have the solutions:
\begin{eqnarray}
a_{101}(t) &=&a_{201}(t)=a_{301}(t)=a_{801}(t)=0 \\
a_{501}(t) &=&C_{1}C_{5}\exp (-\theta _{1}t)+C_{1}C_{6}\exp (-\theta
_{1}t)\\\nonumber
&+&C_{1}C_{7}\exp (-\gamma _{con}t/2) \\
a_{601}(t) &=&C_{1}C_{2}\exp (-\theta _{1}t)+C_{1}C_{3}\exp (-\theta
_{2}t)\\\nonumber
&+&C_{1}C_{4}\exp (-\gamma _{con}t/2)] \\
a_{401}(t) &=&ia_{501}(t) \\
a_{701}(t) &=&-ia_{601}(t)
\end{eqnarray}
where
\begin{eqnarray}
\theta _{1} &=&3\gamma_b /4+\sqrt{\gamma_b }\theta /4,\theta _{2}=3\gamma_b /4-%
\sqrt{\gamma_b }\theta /4 \\
\theta &=&\sqrt{-32\beta ^{2}+\gamma_b } \\
C_{1} &=&\frac{\sqrt{\gamma_b \gamma _{con}}}{\theta \lbrack 2\gamma_b (4\beta
^{2}+\gamma_b )-3\gamma_b \gamma _{con}+\gamma _{con}^{2}]} \\
C_{2} &=&4\sqrt{\gamma_b }(8\beta ^{2}-\gamma_b +\sqrt{\gamma_b }\theta )+\gamma_{con}(\sqrt{\gamma_b }-\theta ) \\
C_{3} &=&4\sqrt{\gamma_b }(-8\beta ^{2}+\gamma_b +\sqrt{\gamma_b }\theta -\gamma_{con}(\sqrt{\gamma_b }+\theta ) \\
C_{4} &=&-4\gamma_b \theta +2\gamma _{con}\theta \\
C_{5} &=&2\sqrt{2}\beta (-3\gamma_b +2\gamma _{con}+\sqrt{\gamma_b }\theta ) \\
C_{6} &=&2\sqrt{2}\beta (3\gamma_b -2\gamma _{con}+\sqrt{\gamma_b }\theta ) \\
C_{7} &=&4\sqrt{2\gamma_b }\beta \theta
\end{eqnarray}
and the density matrix $\rho _{0,1}$ at time $t$ can be presented
as
\begin{equation}
\rho _{0,1}(t)=\frac{1}{2}\bar{a}_{0,1}\bar{\lambda}=\left(
\begin{array}{ccc}
0 & 0 & 0 \\
0 & 0 & 0 \\
a_{401}(t) & a_{601}(t) & 0%
\end{array}%
\right)
\end{equation}
where we used $\gamma _{c}=2\gamma _{b} $ (for a transmon).

The top level equation $\rho _{1,1}(t)$ which represents the actual system evolution is not traceless. In the Bloch representation the equation is  $\rho _{1,1}=\frac{1}{3}\mathrm{I}+\frac{1}{2}\bar{a}_{1,1}\bar{\lambda}$
with initial conditions: $a_{811}(t=0)=-2/\sqrt{3}$ and $a_{i11}(t=0)=0$ for $i\neq 8$. Substituting this expression and the solution for $\rho_{0,1}$ into the master equation we have the motion equation for $\bar{a}_{1,1}$.

Only motion equations for $a_{211}$, $a_{311}$ and $a_{811}$ are coupled and non-zero. We define a vector:
\begin{equation}
x=(a_{211},a_{311,}a_{811})^{T}
\end{equation}%
with initial condition
\begin{equation}
x(0)=(0,0,-\frac{2}{\sqrt{3}})^{T}
\end{equation}
\begin{equation}
\frac{dx}{dt}=Ax+B(t)
\end{equation}%
with coefficient matrices:
\begin{equation}
A=\left(
\begin{array}{ccc}
-3\gamma_b /2 & -2\sqrt{2\gamma_b }\beta & 0 \\
2\sqrt{2\gamma_b }\beta & -5\gamma_b /2 & -\sqrt{3}\gamma_b /2 \\
0 & \sqrt{3}\gamma_b /2 & -\gamma_b /2%
\end{array}%
\right)
\end{equation}
\begin{equation}
B(t)=\left(
\begin{array}{c}
-2\sqrt{\gamma_b }\xi a_{501} \\
-\gamma_b +2\sqrt{\gamma_b }\xi a_{601} \\
-\frac{\gamma_b }{\sqrt{3}}-2\sqrt{3\gamma_b }\xi a_{601}%
\end{array}%
\right)
\end{equation}%
By diagonalization and integration, we obtain the solutions as:
\begin{eqnarray}
&&x[i](t) \\\nonumber
&=&V_{i1}C_{11}(t)[\int [C_{11}(-t^{\prime })Q_{11}B_{1}(t^{\prime
})+C_{11}(-t^{\prime })Q_{12}B_{2}(t^{\prime })\\\nonumber
&+&C_{11}(-t^{\prime
})Q_{13}B_{3}(t^{\prime })]dt^{\prime }-\frac{2}{\sqrt{3}}Q_{13}]  \notag \\\nonumber
&+&V_{12}C_{22}(t)[\int [C_{12}(-t^{\prime })Q_{21}B_{1}(t^{\prime
})+C_{22}(-t^{\prime })Q_{22}B_{2}(t^{\prime })\\\nonumber
&+& C_{22}(-t^{\prime
})Q_{23}B_{3}(t^{\prime })]dt^{\prime }-\frac{2}{\sqrt{3}}Q_{23}]  \notag \\\nonumber
&+&V_{13}C_{33}(t)[\int [C_{13}(-t^{\prime })Q_{31}B_{1}(t^{\prime
})+C_{33}(-t^{\prime })Q_{32}B_{2}(t^{\prime })\\\nonumber
&+&C_{33}(-t^{\prime
})Q_{33}B_{3}(t^{\prime })]dt^{\prime }-\frac{2}{\sqrt{3}}Q_{33}]  \notag \\\nonumber
&=&V_{i1}S_{1}+V_{i2}S_{2}+V_{i3}S_{3}-\frac{2}{\sqrt{3}}%
V_{i1}C_{11}(t)Q_{13}\\\nonumber
&-&\frac{2}{\sqrt{3}}V_{i2}C_{22}(t)Q_{23}-\frac{2}{\sqrt{%
3}}V_{i3}C_{33}(t)Q_{33}
\end{eqnarray}

where%
\begin{eqnarray}
S_{i} &=&C_{1i}(t)\int [C_{1i}(-t^{\prime })Q_{i1}B_{1}(t^{\prime
})+C_{1i}(-t^{\prime })Q_{i2}B_{2}(t^{\prime })\\\nonumber
&+&C_{1i}(-t^{\prime
})Q_{i3}B_{3}(t^{\prime })]dt^{\prime } \\\nonumber
&=&\frac{\gamma_b }{\lambda _{i}}\left( Q_{i2}+\frac{1}{\sqrt{3}}Q_{i3}\right)
\left( 1-\exp \left( \lambda _{i}t\right) \right) \\\nonumber
&+&\frac{2\sqrt{\gamma_b \gamma _{con}}C_{1}C_{4}}{\gamma _{con}+\lambda _{i}}%
( Q_{i2}-\sqrt{3}Q_{i3}) ( \exp (-(\gamma_b +\gamma_{con})t)\\\nonumber
&-&\exp ( \lambda _{i}t))  \notag \\\nonumber
&-&\frac{2\sqrt{\gamma_b \gamma _{con}}C_{1}C_{2}}{\Delta _{1}+\lambda
_{i}+\frac{\gamma _{con}}{2}}\left( Q_{i2}-\sqrt{3}Q_{i3}\right)( \exp (-(\Delta
_{1}+\frac{\gamma _{con}}{2})t)\\\nonumber
&-&\exp( \lambda _{i}t))  \notag \\\nonumber
&-&\frac{2\sqrt{\gamma_b \gamma _{con}}C_{1}C_{3}}{\Delta _{2}+\lambda
_{i}+\frac{\gamma _{con}}{2}}\left( Q_{i2}-\sqrt{3}Q_{i3}\right)( \exp (-(\Delta
_{2}+\frac{\gamma _{con}}{2})t)\\\nonumber
&-&\exp ( \lambda _{i}t))  \notag
\end{eqnarray}
with matrices $V$, $Q$ and $C$ being the eigen-vector matrix, inverse eigen-vector matrix and the diagonalized matrix of the coefficient matrix $A$.

Then we have the system density matrix at time t:
\begin{eqnarray}
&&\rho _{1,1}(t)=\frac{1}{3}\mathrm{I}\\\nonumber
&+&\frac{1}{2}\left(
\begin{array}{ccc}
a_{311}(t)+\frac{a_{811}(t)}{\sqrt{3}} & -ia_{211}(t) & 0 \\
ia_{211}(t) & \frac{a_{811}(t)}{\sqrt{3}}-a_{311}(t) & 0 \\
0 & 0 & -2\frac{a_{811}(t)}{\sqrt{3}}%
\end{array}%
\right) ,
\end{eqnarray}
The transmon polarisation for the homodyne detection is
\begin{equation}
\left\langle \hat{y}(t)\right\rangle=\rm Tr[-i(\hat{L}_{c}-\hat{L}_{c}^{\dag })\rho _{1,1}](t)=\rm Tr\left[ \lambda _{2}\rho _{1,1}%
\right](t) =a_{211}(t)
\end{equation}
\begin{figure}[tbp]
\centering
\includegraphics[ width = 0.45\textwidth]{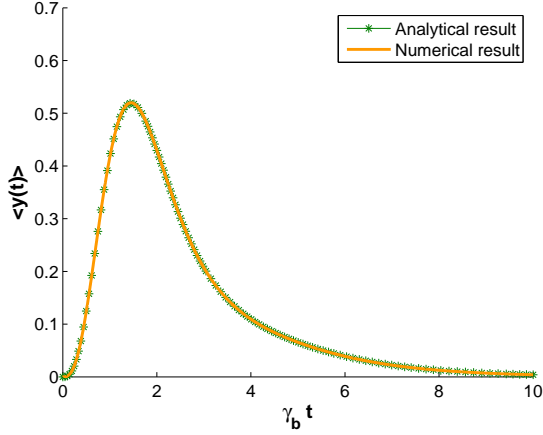}
\caption{(Color online) The time evolution of the transmon polarisation: comparison between analytical and numerical results. The parameters are: $\gamma_c=2\gamma_b, \gamma_{con}=\gamma_b, \Delta_c=\Delta_b=0, \beta=\gamma_b$.}
\label{y_AN}
\end{figure}
In Fig. \ref{y_AN}, we compare the results for $\left\langle \hat{y}(t)\right\rangle$ by analytical and numerical methods. It is a perfect agreement between the results by the two methods.

The noise or the variance of the detected signal, can be calculated by the quantum regression theorem \cite{lax63}:
\begin{eqnarray}
&&(\Delta S)^{2}=E\left[ S^{2}-\bar{S}^{2}\right]  \\\nonumber
&&=\int^{T}_0 dt\int^{T}_0 dt' u(t'-t)\rm Tr[\hat{y}(t)e^{\mathcal{L}(t'-t)}(-i\hat{L}_c\rho(t)\\\nonumber
&&+i\rho(t)\hat{L}^\dag_c)]+u(t-t')\rm Tr[\hat{y}e^{\mathcal(t-t')}(-i\hat{L}_c\rho(t')\\\nonumber
&&+i\rho(t')\hat{L}^\dag_c)]+T-\bar{S}^{2}
\end{eqnarray}%
where $S=\int^T_0 dtJ_{hom}(t)$, $\bar{S}=\int^T_0 dt y_{uc}(t)$ and the function $u(t)=1 (t>0); u(t)=0 (t<0)$. The subscripts $uc$ means unconditional results.

\subsection{Squeezed probe}
\label{squeezing}

In this subsection we replace the coherent probe field of a phase-squeezed state and the corresponding stochastic master equation is

\begin{eqnarray}
 &&d\rho=
(-i[H_s,\rho]+\gamma_{con}\mathcal{D}[\hat{a}_{con}]\rho+\gamma_b\mathcal{D}[\hat{\sigma}_{ab}]\rho
\\\nonumber
&-&\sqrt{\gamma_{con}\gamma_b}([\hat{\sigma}_{ba},\hat{a}_{con}\rho]+[\rho\hat{a}^\dag_{con},\hat{\sigma}_{ab}])+\gamma_c (N+1)\mathcal{D}[\hat{\sigma}_{bc}]\\\nonumber
&+&\gamma_c
N\mathcal{D}[\hat{\sigma}_{cb}]+\gamma_c
M\hat{\sigma}_{bc}\rho\hat{\sigma}_{bc}-\gamma_c
M^*\hat{\sigma}_{cb}\rho\hat{\sigma}_{cb})dt\\\nonumber
&+&\sqrt{\frac{\gamma_c}{L}}\mathcal{H}[(N+1+M)\hat{\sigma}_{bc}e^{-i\pi/2}-(N+M^*)\hat{\sigma}_{cb}e^{i\pi/2}]\rho
dW \label{SME}
\end{eqnarray}
where
\begin{eqnarray}
H_s &=&
\Delta_{c}\hat{\sigma}_{cc}+\Delta_{b}\hat{\sigma}_{bb}+\sqrt{\gamma_c}\int d\nu
\beta_\nu(\hat{\sigma}_{bc}+\hat{\sigma}_{cb})
\end{eqnarray}
with $M=\sinh(r)\cosh(r)e^{i\theta}$, $N=\sinh^2(r)$ and $L=1+2N+M+M^*$. The instantaneous photocurrent is
\begin{eqnarray}
I^{hom}_c(t)=\left\langle
\hat{y}\right\rangle_c(t)+\sqrt{L}\xi(t)
\end{eqnarray}
This equation indicates that the noise term $\xi$ is multiplied by a factor of $\sqrt{L}$. Notice that when $\theta$ is chosen as
zero, $L=e^{-2r}$ and the noise is reduced while when $\theta=\pi$, the noise is amplified.

\begin{figure}[Htbp]
\centering
\includegraphics[ width =0.45\textwidth]{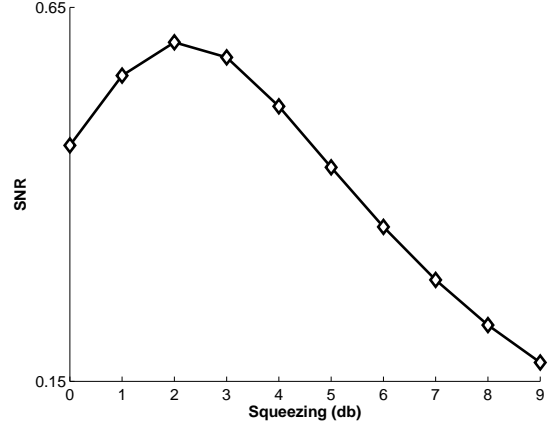} \caption{{\footnotesize The SNR as a function of the squeezing degree (db). The parameters are: $\gamma_c=2\gamma_b, \gamma_{con}=0.6672\gamma_b, \beta=0.4\gamma_b, \Delta_c=\Delta_b=0$.}}\label{SNR_db}
\end{figure}

In Fig. \ref{SNR_db} we show that the squeezing in the phase quadrature can only help to improve the SNR slightly. According to uncertainty relation the phase squeezing indicates an amplification of the amplitude noise, which results in larger dynamical noise in the atomic response.

\subsection{Cascaded n transmons}
In the main text we have discussed the single transmon and a transmon ensemble for single microwave detection. In this subsection, we investigate the feasibility of cascading multiple transmons, each with a probe field.

\begin{figure}[Htbp]
\centering
\includegraphics[ width = 0.45\textwidth]{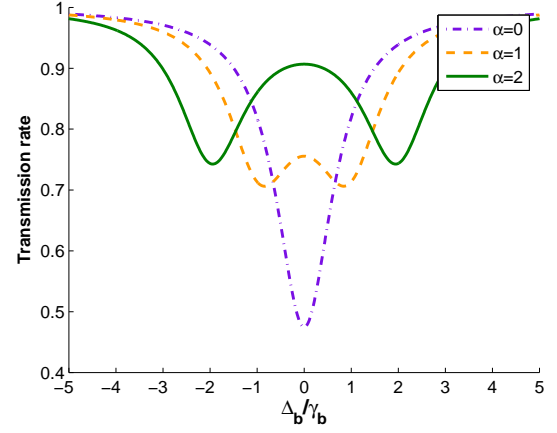} \caption{{\footnotesize (Color online) Transparency spectrum.}}\label{T_D1}
\end{figure}

First we evaluate the transmission rate of the signal field after passing one transmon. As is known that in low dimensional systems photons will be
reflected completely by qubits on resonance \cite{SHFan05}\cite{zumofen}. On the other hand, EIT-like effects appear when there are more energy levels \cite{SHFan09}. In this three-level structure, two-pathway interference forms and the familiar EIT-like transparency window appears, shown in Fig. \ref{T_D1}. However, how wide is the window and whether the parameters for this transparency window is consistent with those at best SNR are questions. In the following we calculate the transmission rate by the similar procedure in \cite{SHFan05}.

The cascaded system is equivalent to the model of transmons
interacting with a single photon pulse centered at $\omega_c$ and a
coherent field in the transmission line directly. After linearizing the dispersion in the vicinity of $\omega_c$ and formally adding the non-Hermitian damping terms (which come from the Markovian approximation by tracing out the bath operators), the Hamiltonian in the real space can be written as
\begin{eqnarray}
H &=& \int
dx[\hat{a}^\dag_R(x)(\omega_c-iv_g\frac{\partial}{\partial
x})\hat{a}_R(x)\\\nonumber
&+&\hat{a}^\dag_L(x)(\omega_c+iv_g\frac{\partial}{\partial
x})\hat{a}_L(x)]+(\omega_{c}-\omega_p-i\gamma_c/2)\hat{\sigma}_{cc}\\\nonumber
&+&(\omega_b-i\gamma_b/2)\hat{\sigma}_{bb}+\int
dx\sqrt{\gamma_b}\delta(x)[\hat{a}^\dag_R\hat{\sigma}_{ab}+\hat{\sigma}_{ba}\hat{a}_R\\\nonumber
&+&\hat{a}^\dag_L\hat{\sigma}_{ab}+\hat{\sigma}_{ba}\hat{a}_L]+\sqrt{\gamma_c}\alpha(\hat{\sigma}_{bc}+\hat{\sigma}_{cb})
\end{eqnarray}
where $v_g$ is the group velocity of the signal photon, which depends on the geometry and
material of the waveguide. For a typical coplanar waveguide of
quantum circuit system, $v_g=\frac{1}{\sqrt{C'L'}}=\frac{1}{C'
Z_0}=c/\sqrt{\epsilon_{eff}}$ with the effective permittivity
$\epsilon_{eff}$ around 5.9. The time-independent eigen-equation is
\begin{eqnarray}
H\left\vert E_k\right\rangle=E_k\left\vert E_k\right\rangle
\end{eqnarray}
with $E_k=\omega=\omega_c+v_g k_R$ and
\begin{eqnarray}
\left\vert E_k\right\rangle&=&\int dx\phi_R\hat{a}^\dag_R(x)\left\vert 0_R,0_L,a\right\rangle\\\nonumber
&+&\int dx \phi_L\hat{a}^\dag_L\left\vert0_R,0_L,a\right\rangle\\\nonumber
&+&c_1\left\vert0_R,0_L,b\right\rangle+c_2\left\vert 0_R,0_L,c\right\rangle
\end{eqnarray}

Then we have the equations for the coefficients:
\begin{eqnarray}
(\omega_{con}-iv_g)\frac{\partial}{\partial x}\phi_R+g_1\delta(x)c_1=\omega\phi_R
\end{eqnarray}
\begin{eqnarray}
(\omega_{con}+iv_g)\frac{\partial}{\partial
x}\phi_L+\alpha\delta(x)c_1=\omega\phi_L
\end{eqnarray}
\begin{eqnarray}
(\omega_b-i\frac{\gamma_c}{2})c_1+\sqrt{\gamma_b}(\phi_R(0)+\phi_L(0))+\sqrt{\gamma_c}c_2\alpha=\omega c_1
\end{eqnarray}
\begin{eqnarray}
(\omega_c-i\gamma_c/2-\omega_p)c_2+\sqrt{\gamma_c}c_1\alpha=\omega c_2
\end{eqnarray}
where
\begin{eqnarray}
\phi_R=exp(ikx)\theta(-x)+t exp(ikx)\theta(x)
\end{eqnarray}
\begin{eqnarray}
\phi_L=rexp(-ikx)\theta(-x)
\end{eqnarray}
with $\theta(x)$ being the Heaviside step function.

By solving the equations above, the transmission amplitude can be
obtained;

\begin{eqnarray}
t=\frac{(\Delta'_b+i\sqrt{\gamma_c}/2)(\Delta'_c+i\gamma_c/2)-\gamma_c\alpha^2}{(\Delta'_b+i\gamma_c/2+i
\gamma_b/vg)(\Delta'_c+i\gamma_c/2)-\gamma_c\alpha^2}
\end{eqnarray}
where $\Delta_b'=\omega_b-\omega$ and $\Delta_c'=\omega_c-\omega_p-\omega$. Here the frequency width of the signal photon is much smaller than its central frequency, that is to say, it is a relatively narrow pulse, therefore in the the following we use $\Delta_b$ and $\Delta_c$ to replace $\Delta_b'$ and $\Delta_c'$.

From Fig. \ref{T_D1}, an induced transparency window appears in our system as expected, which is caused by the two-channel interference between transmon transitions. The width of window is twice of the coupling $\sqrt{\gamma_c}\alpha$. One can find that to achieve a high transmission rate, either large $\alpha$ or large $\Delta_b$ is required. This indicates that a large transmission rate corresponds a low SNR.

\begin{figure}
    \centering
    \includegraphics[width=0.45\textwidth]{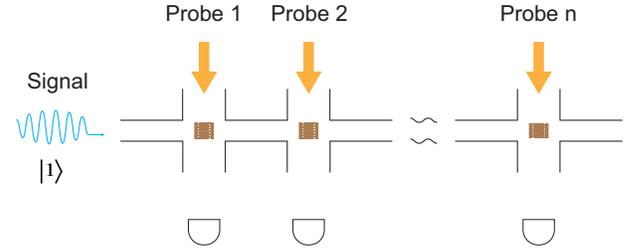}
    \caption{(Color online) The schematic for microwave photon counting using cascaded n tranmons.}
    \label{scheme_n}
  \end{figure}

\begin{figure}
    \centering
    \includegraphics[width=0.45\textwidth]{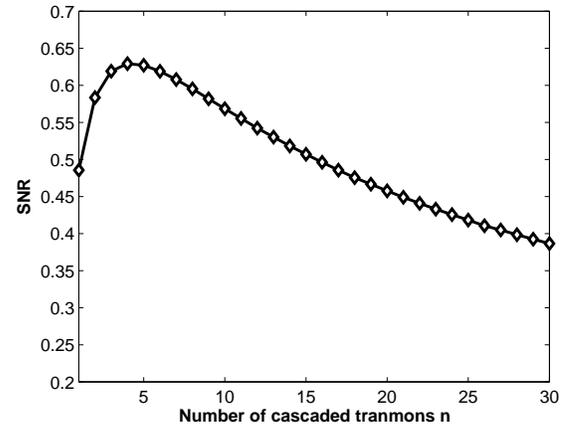}
    \caption{The plot of SNR as a function of transmon number n.}
    \label{n_probe}
\end{figure}

Assuming we cascade n transmons with separate probes and detectors, seen in Fig. \ref{scheme_n} ideally it is equivalent to average over n trajectories. However, after including the reflection, the effective $SNR_n$ becomes (high order terms $o(R)$ are omitted here):
\begin{eqnarray}
SNR_{n}=SNR_{1}(\sqrt{n}T^{n-1}+\sum^{n-1}_{j=1}j/\sqrt{n}T^{j-1}R)
\end{eqnarray}
where $SNR_1$ is the SNR in one transmon and one probe case. After optimizing parameters, we obtain the SNR as a function of probe number as shown in Fig. \ref{n_probe}. Obviously, the signal can not win the noise in this method.

\subsection{Conversion from N-type Four-level structure to ladder-type three-level structure}

The N-type four-level structure has been suggested to be a promising candidate for implementing cross-Kerr nonlinearity. Here in this subsection, we show that it can be approximately mapped to a three-level ladder system.

The Hamiltonian for a four-level system coupling with a signal field $\beta$ at transition 0-1, a control field $\beta$ at transition 1-2 and a probe $\alpha$ at transition 2-3 in a rotating frame is given by
\begin{eqnarray}
H&=&(\Delta_{10}-\Delta_{12})\hat{\sigma}_{00}+\Delta_{12}\hat{\sigma}_{11}+\Delta_{32}\hat{\sigma}_{33}\\\nonumber
&+& \sqrt{\gamma_{01}}\beta(\hat{\sigma}_{10}+\hat{\sigma}_{01})+\frac{\Omega}{2}(\hat{\sigma}_{12}+\hat{\sigma}_{21})\\\nonumber
&+&\sqrt{\gamma}_{32}\alpha(\hat{\sigma}_{23}+\hat{\sigma}_{32})
\end{eqnarray}
When the transition 1-2 is strongly driven by $\Omega$, the subsystem of the transition 1-2 and the control field can be diagonalized independently. We define a unitary transformation U as
\begin{equation}
\left(
\begin{array}{ccc}
 \cos\theta & -\sin\theta  \\
 \sin\theta & \cos\theta
\end{array}
\right)
\end{equation}

By diagonalizing $U^\dag H_{12,\Omega} U$, we have:
\begin{eqnarray}
\theta = \frac{1}{2}\arctan(\Omega/\Delta_{12})
\end{eqnarray}%
and the states ${\left\vert 1\right\rangle, \left\vert 2\right\rangle}$ can be represented in the dressed states $\left\vert -\right\rangle$ and $\left\vert +\right\rangle$\cite{Cohen}:
\begin{eqnarray}
{\left\vert 1\right\rangle}=\cos\theta\left\vert - \right\rangle-\sin\theta\left\vert + \right\rangle\\\nonumber
{\left\vert 2 \right\rangle}=\cos\theta\left\vert +\right\rangle+\sin\theta\left\vert- \right\rangle
\end{eqnarray}
Then the Hamiltonian can be rewritten as
\begin{eqnarray}
H&=&-\Delta_{10}\hat{\sigma}_{00}+\Delta_{32}\hat{\sigma}_{33}+\lambda_+\hat{\sigma}_{++}+\lambda_-\hat{\sigma}_{--}\\\nonumber
&+& \sqrt{\gamma_{01}}\beta(\cos\theta(\hat{\sigma}_{-0}+\hat{\sigma}_{0-})-\sin\theta(\hat{\sigma}_{+0}+\hat{\sigma}_{0+}))\\\nonumber
&+&\sqrt{\gamma}_{32}\alpha(\cos\theta(\hat{\sigma}_{+3}+\hat{\sigma}_{3+})+\sin\theta(\hat{\sigma}_{-3}+\hat{\sigma}_{3-}))
\end{eqnarray}
where $\lambda_{\pm}=\frac{1}{2}\Delta_{12}\pm \frac{1}{2}\sqrt{\Delta^2_{12}+\Omega^2}$.
%\begin{figure}[tbHp]
%\includegraphics[width = 3.5in]{4LS.eps}
%\caption{Conversion from an N-type four-level system to a three-level ladder system }.
%\end{figure}
%\begin{figure}[tbHp]
%\includegraphics[width = 3.5in]{N4L_ladder3L.eps}
%\caption{\footnotesize (Color online) The comparison of the population of the excited state in the two configurations. The parameters are $\Delta_{12}=0$, %$\Delta_{10}=\Delta_{32}=\lambda_-$, $\Omega=10\gamma_{10}$, and $\alpha=\beta=\gamma_{12}=\gamma_{32}=\gamma_{10}$.}
%\label{4L_3L}
%\end{figure}

\begin{figure}
    \centering
    \includegraphics[ width = 0.45
\textwidth]{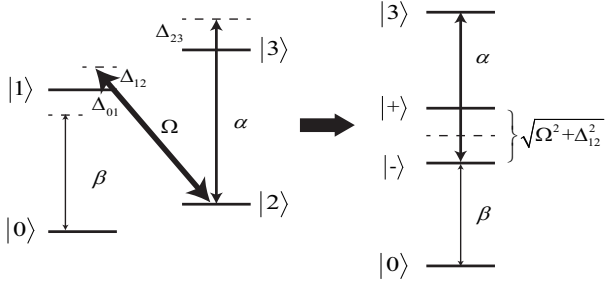}
    \caption{Illustration of the conversion from an N-type four-level system to a ladder-type three-level system.}
\end{figure}
\begin{figure}
    \centering
    \includegraphics[width=0.45\textwidth]{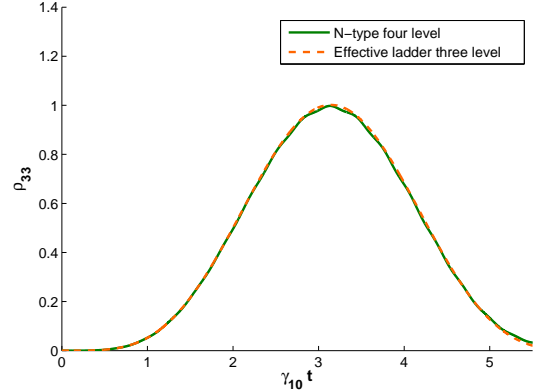}
    \caption{(Color online)  The comparison of the population of the excited state in the two configurations. The parameters are $\Delta_{12}=0$, $\Delta_{10}=\Delta_{32}=\lambda_-$, $\Omega=10\gamma_{10}$, and $\alpha=\beta=\gamma_{12}=\gamma_{32}=\gamma_{10}$.}
    \label{4L_3L}
\end{figure}

We assume the signal field and probe field are tuned to be resonant with the state $\left\vert -\right\rangle$, that is, $\Delta_{10}=\lambda_-$ and $\Delta_{32}=\lambda_-$. We perform a rotating and ignore the fast-varying terms like $e^{i\sqrt{\Delta^2_{12}+\Omega^2}t}$, then we have an approximate three level system:
\begin{eqnarray}
H= \cos\theta\sqrt{\gamma_{01}}\beta(\hat{\sigma}_{-0}+\hat{\sigma}_{0-})+\sin\theta\sqrt{\gamma_{32}}\alpha(\hat{\sigma}_{-3}+\hat{\sigma}_{3-})
\end{eqnarray}
As shown in Fig. \ref{4L_3L}, it is a good approximation when there is a strong driving ($\Omega\gg \gamma_{01}, \alpha, \beta$) at the dressed transition.
%\end{multicols*}
\subsection{Effect of different relaxation rate ratio $\gamma_c/\gamma_b$}

\begin{figure}[Htbp]
\centering
\includegraphics[ width = 0.45
\textwidth]{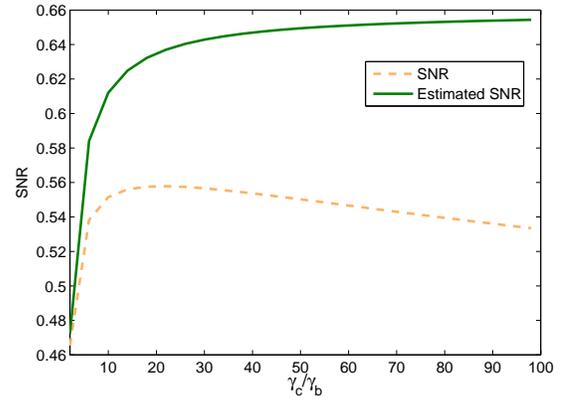} \caption{{\footnotesize (Color online) The SNR as a function of the ratio $\gamma_c/\gamma_b$.}}\label{rc2rb}
\end{figure}

For a transmon, the relaxation rate of the upper transition is twice that of the lower transition but in other three-level system we may have different ratio of the relaxation rates. Here we investigate the effect of different relative relaxation rates on the SNR numerically. Fig. \ref{rc2rb} shows the SNR as a function of the ratio $\gamma_c/\gamma_b$. It is clear that no matter lower ratio or higher ratio the best SNR can not be improved much and still far below unity.

\bibliography{xuan_ref}
\end{document}